\documentclass[british]{article}\usepackage[]{graphicx}\usepackage[]{xcolor}
\makeatletter
\def\maxwidth{ %
  \ifdim\Gin@nat@width>\linewidth
    \linewidth
  \else
    \Gin@nat@width
  \fi
}
\makeatother

\definecolor{fgcolor}{rgb}{0.345, 0.345, 0.345}

\usepackage{framed}
\makeatletter
 {\par\unskip\endMakeFramed%
 \at@end@of@kframe}
\makeatother

\definecolor{shadecolor}{rgb}{.97, .97, .97}
\definecolor{messagecolor}{rgb}{0, 0, 0}
\definecolor{warningcolor}{rgb}{1, 0, 1}
\definecolor{errorcolor}{rgb}{1, 0, 0}
\newenvironment{knitrout}{}{} 

\usepackage{alltt}
\PassOptionsToPackage{natbib=true}{biblatex}

\usepackage[T1]{fontenc}
\usepackage[latin9]{inputenc}
\usepackage{geometry}
\usepackage{cprotect}
\usepackage{url}
\usepackage{bm}
\usepackage{amsmath}
\usepackage{amssymb}
\usepackage{rotfloat}
\usepackage{setspace}

\makeatletter
\newcommand{\lyxaddress}[1]{
	\par {\raggedright #1
	\vspace{1.4em}
	\noindent\par}
}

\makeatother

\usepackage{babel}
\usepackage[style=authoryear, isbn=false, url=false, uniquename=false, uniquelist=false, maxbibnames=50, date=year, giveninits=true]{biblatex}
\IfFileExists{upquote.sty}{\usepackage{upquote}}{}
\begin{document}

\title{Bayesian views of generalized additive modelling}
\author{David L. Miller}
\maketitle

\lyxaddress{Biomathematics and Statistics Scotland, Dundee, Scotland. }

\lyxaddress{UK Centre for Ecology \& Hydrology, Lancaster Environment Centre,
Lancaster, United Kingdom}

\begin{enumerate}
\item Generalized additive models (GAMs) are a frequently used, flexible
framework applied to many problems in statistical ecology. They are
commonly used to incorporate smooth effects into models via splines,
including spatial components in species distribution models.
\item GAMs are often considered to be a purely frequentist framework (`generalized
linear models with wiggly bits'), however links between frequentist
and Bayesian approaches to these models were highlighted early-on
in the literature. From a practical perspective, Bayesian thinking
underlies many parts of the implementation in the popular R package
\texttt{mgcv}, so understanding these underpinnings can be informative
during model building and assessment. 
\item This article aims to highlight useful links (and differences) between
Bayesian and frequentist approaches to smoothing, as detailed in the
statistical literature, in accessible way, with a focus on the \texttt{mgcv}
implementation. By harnessing these links we can expand the set of
modelling tools we have at our disposal, as well as our understanding
of how existing methods work.
\item Two important topics for quantitative ecologists are covered in detail:
model term selection and uncertainty estimation. Taking Bayesian viewpoints
for these problems makes them much more tractable in many applied
settings. Examples are given using data from the NOAA Alaska Fisheries
Science Center's groundfish assessment program.
\end{enumerate}

\section{Introduction}

Flexible modelling of responses for a variety of distributions (binary,
count, bounded, continuous) is an indispensable tool for quantitative
ecologists. Common applications include species distribution modelling
\citep{golding_fast_2016}, abundance estimation \citep{miller_estimating_2022},
dose-response modelling \citep{jacobson_quantifying_2022}, movement
\citep{aarts_comparative_2012}, ecosystem health \citep{augustin_modeling_2009}
and more. In each case what is important is incorporating the structure
of the data and/or data collection process into the model (be that
the form of relationships, spatial correlation, blocking effects,
etc). Informally this structure can be thought of as imposing some
prior on how we would like the terms in the model to behave. In this
article, I regurgitate some results from the statistical literature
emphasizing this (Section \ref{sec:Bayesian-interpretations}) and
then show how these tools can be used (or are already used) by those
engaged in ecological modelling (Section \ref{sec:Some-examples}).

Generalized Additive Models \citep[GAMs; e.g.,][]{wood_generalized_2017-1}
are often taught as an extension of the linear model: adding wiggles
(via smoothers) to make a (G)LM more flexible, often as a more principled
step forward from adding polynomial terms. ``Smooth'' is often a
synonym for spline \citep{deboor_practical_1978}, but there are many
possible model terms that can be specified as \textit{basis functions}
subject to \textit{penalties}: so-called ``basis-penalty smoothers''.
This class of models includes ranges from very simple structures (random
effects), through to more structured penalties, obtaining more complex
hierarchical random effects models. Having the penalty encode spatial
information about a graphical structure gives (Gaussian) Markov random
fields \citep{rue_gaussian_2005} or multivariate spline models like
thin-plate regression splines \citep{wood_thin_2003}. We can also
use tensor products of terms to construct multidimensional interaction-type
effects \citep{wood_modelling_2000}, allowing for different units
to be used for each covariate (i.e., anisotropy). Here I use the word
smooth to include all these possible flexible model terms and generally
denote them as $s()$.

Though the term ``GAM'' has significant baggage regarding the fitting
method and type of terms, it really just describes the form of the
linear predictor in the model (terms add together) and the response
distribution. For example, a model may look like:

\begin{equation}
g(\mu_{i})=\bm{a}_{i}^{\intercal}\bm{\theta}+s_{1}(x_{1i})+s_{2}(x_{2i})+s_{3}(x_{3i},x_{4i})\label{eq:gam}
\end{equation}
where $\mu_{i}\equiv\mathbb{E}(Y_{i})$ and $Y_{i}\sim\text{EF}(\mu_{i},\phi)$
where $Y_{i}$ ($i=1,\ldots,n$) is the response and $\text{EF}(\mu_{i},\phi)$
indicates an exponential family distribution with mean $\mu_{i}$
and scale parameter $\phi$. $\bm{a}_{i}^{\intercal}$ is a vector
of slopes and intercept covariates, where $\bm{\theta}$ are their
associated coefficients. The $s_{j}$ are ``smooth'' functions of
one or more of the covariates $x_{1i},x_{2i},x_{3i},x_{4i}$. This
definition can be adapted to Generalized Additive Mixed Models (GAMMs)
and Generalized Linear Mixed Models (GLMMs), as we will see below.

\begin{figure}
\begin{knitrout}
\definecolor{shadecolor}{rgb}{0.969, 0.969, 0.969}\color{fgcolor}
\includegraphics[width=\maxwidth]{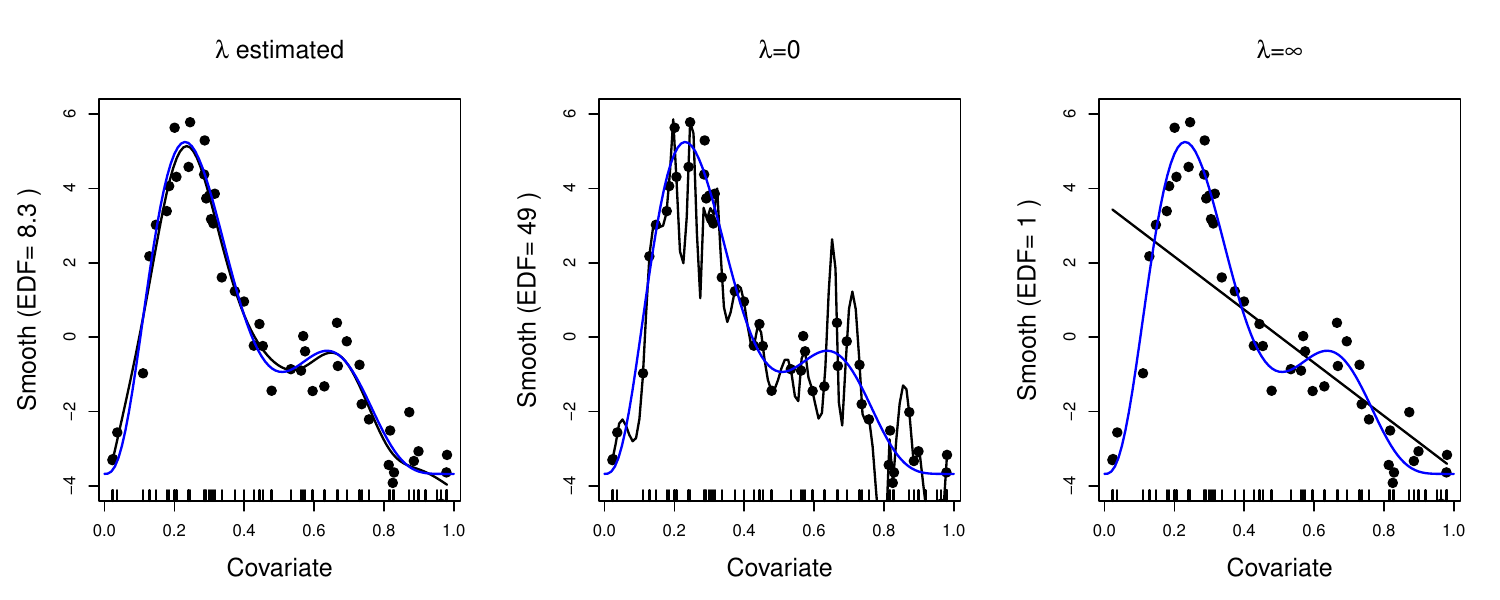} 
\end{knitrout}

\caption{The effect of smoothing parameters on the effective degrees of freedom
of a smooth. In each case, data was simulated from the true, blue,
function with normal noise (with zero mean and standard deviation
of 0.5) added. The data are shown as points. In each of the three
plots, a thin-plate regression spline was fitted to the data with
differing smoothing parameters. In the right plot, the smoothing parameter,
$\lambda$, was estimated from the data, giving an effective degrees
of freedom (EDF) of 8.3. In the middle plot the smoothing parameter
was set to zero, meaning the penalty has no effect, leading to a very
wiggly fit (EDF is the maximum). In the left plot the smoothing parameter
was set to be (numerically) infinite, leading to a penalty that doesn't
allow for any wiggles. This leads to an EDF of 1, leaving only a linear
fit (since this has no second derivative, it lies in the nullspace
of the penalty and is unpenalized; see Section \ref{subsec:Term-selection,-proper}).\protect\label{fig:EDF-lambda}}

\end{figure}

The smooth terms are what makes GAMs an interesting and useful evolution
of the generalized linear model. In a very general sense, they are
constructed from sums of simple basis functions \citep[e.g.,][]{deboor_practical_1978}.
We can construct a complicated function by summing smaller, less complicated
\textit{basis functions}. In general for some smooth $s$ of covariate
$x$ we have the following decomposition:

\begin{equation}
s(x)=\sum_{k=1}^{K}\beta_{k}b_{k}(x),\label{eq:basis}
\end{equation}
where $b_{k}$ are fixed basis functions (with maximum complexity
or basis dimension $K$) and $\beta_{k}$ are coefficients to be estimated.
This basis function approach is extremely flexible, so to avoid overfitting
we penalize the flexibility of each smooth term according to its \textit{wiggliness.}
This means that we can let $K$ be relatively large, and let the penalty
remove the extra flexibility. The fitted model has much smaller \textit{effective
degrees of freedom} (EDF); that is, the degrees of freedom actually
used by the model, once the penalty is taken into account (usually
defined as the sum of the diagonal elements of the hat matrix; \citet[Section 5.4.2]{wood_generalized_2017-1}).
Figure \ref{fig:EDF-lambda} illustrates these concepts. Generally
such a penalty will be an integral (sometimes a sum) of squared derivatives
of $s$ (since derivatives measure the changes in $s$). The penalty
can be written in the form $\sum_{m=1}^{M}\lambda_{m}\bm{\beta}^{\intercal}\bm{S}_{m}\bm{\beta}$
where $\bm{\beta}$ is a vector of coefficients, $\bm{S}_{m}$ is
a matrix of the fixed parts of the penalty (integrated, squared derivatives
of the $b_{k}$s, which do not change) and $\lambda_{m}$ are smoothing
parameters to be estimated that control the influence of the penalty
(where the $\bm{S}_{m}$s are padded with zeros so the sum forms a
block matrix); see e.g., \citet[Section 4.2.2]{wood_generalized_2017-1}.
Writing the penalty in this way means that we can compute the $\bm{S}_{m}$s
once and during fitting the penalty is calculated by matrix multiplication
only. Note that multiple penalty terms can correspond to a single
smooth or multiple smooths may share a single smoothing parameter
so $M$ is not necessarily the number of smooth terms in the model.

We want to estimate model parameters that describe the data best,
in the sense that we want to draw lines (or surfaces etc) that are
close to the data but do not interpolate them. Formally we can set
this up as a penalized log-likelihood \citep{hastie_bayesian_2000}
to find:

\begin{equation}
\hat{\bm{\beta}}=\underset{\bm{\beta}}{\text{argmax}}\left\{ l(\bm{\beta})-\sum_{m=1}^{M}\lambda_{m}\bm{\beta}^{\intercal}\bm{S}_{m}\bm{\beta}\right\} ,\label{eq:betaarg}
\end{equation}
where $l$ is the log-likelihood and there are $M$ smoothing parameters
to estimate. Here we are trading-off between fit (high log-likelihood
giving smooths close to the data) and penalty (large penalty indicating
smooths are too wiggly).

Conditional on the $\lambda_{m}$s, estimation of $\hat{\bm{\beta}}$
in (\ref{eq:betaarg}) is relatively straight-forward and the problem
can be attacked with penalized iteratively re-weighted least squares
(PIRLS) as for a GLM \citep[section 6.1.1]{wood_generalized_2017-1}.
Estimating both $\hat{\bm{\beta}}$ and $\hat{\bm{\lambda}}$ is more
complicated, as the smoothing parameters will constrain the values
that $\hat{\bm{\beta}}$ can take. There have been various proposals
for methods to fit such models in a frequentist framework and these
fall into two categories: prediction error minimizing methods, like
generalized cross-validation (GCV) or Akaike's information criterion
(AIC), or likelihood based methods like restricted maximum likelihood
and marginal likelihood (REML/ML) \citep[see][for a review]{wood_fast_2011}.
Prediction error minimizing methods have been shown to overfit (undersmooth)
at finite sample sizes \citep[for GCV;][]{reiss_smoothing_2009},
especially when errors are correlated \citep[for AIC;][]{krivobokova_note_2007},
so REML/ML have become the preferred methods. REML and ML cast the
smooth functions as random effects \citep{ruppert_semiparametric_2003}
and smoothing parameters as variance parameters, so we can think of
$\boldsymbol{\beta}$ as being a multivariate normal random effect
with a variance proportional to $\boldsymbol{\lambda}$ (with structure
imposed by the penalty matrix/matrices). 

When we talk about adding smooth functions to our model, we tend to
concentrate on equations like (\ref{eq:gam}), looking at the mean
effects of including smooths rather than thinking about the penalty.
We usually view the penalty as a way of constraining our fit, stopping
it from being too wiggly and ensuring that our model does not overfit.
We can also think of the basis-penalty as the consequence of the problem
definition, we have chosen them due to what we know about the dependencies
and structures in the data (or data collection process). In practice,
for univariate smoothing, switching between basis functions does not
tend to make a big impact on results unless there are clear features
that need to be accounted for (such as cyclic phenomena, boundary
issues etc); it is certainly not the case that one should spend time
searching for an `optimal' basis. In a loose sense selecting the
basis is equivalent setting-up a prior on the kinds of functions we
want to fit. The rest of this article investigates this idea further
and explores some useful applications in ecology.

\section{Bayesian interpretations\protect\label{sec:Bayesian-interpretations}}

We can quickly get to a convenient Bayesian formulation by exponentiating
the objective function in (\ref{eq:betaarg}) \citep[section 5.8]{wood_generalized_2017-1},
which in the frequentist case gives us a penalized likelihood $\mathcal{L}_{p}$:
\begin{equation}
\mathcal{L}_{p}(\boldsymbol{\beta},\boldsymbol{\lambda})=\mathcal{L}(\boldsymbol{\beta})\exp{\left(-\boldsymbol{\beta}^{\intercal}\mathbf{S}_{\boldsymbol{\lambda}}\boldsymbol{\beta}\right)}.\label{eq:bayes-lik}
\end{equation}
We recognise this as Bayes theorem: we might better write $\mathcal{L}_{p}(\boldsymbol{\beta},\boldsymbol{\lambda})$
as $p(\boldsymbol{\beta}\vert\boldsymbol{\lambda},\mathbf{y})$ (the
posterior for $\boldsymbol{\beta}$) and the likelihood $\mathcal{L}(\boldsymbol{\beta})$
as $p(\mathbf{y}\vert\boldsymbol{\lambda},\boldsymbol{\beta})$. Finally,
$\exp{\left(-\boldsymbol{\beta}^{\intercal}\mathbf{S}_{\boldsymbol{\lambda}}\boldsymbol{\beta}\right)}$
acts as a prior on $\boldsymbol{\beta}$, $p(\boldsymbol{\beta})$.
This prior is proportional to a multivariate normal distribution with
mean zero and think of $\mathbf{S}_{\boldsymbol{\lambda}}$ (defined
as $\mathbf{S}_{\lambda}=\sum_{m}\lambda_{m}\mathbf{S}_{m}$) as a
prior precision matrix. 

\subsection{Specifying priors\protect\label{subsec:Priors}}

By using a smooth term for a given covariate in our model we are specifying
that observations which are close to each other (in some sense) in
covariate space have similar values, the response varies smoothly
(according to some measure of smoothness) and that the true function
we seek to estimate is more likely to be smooth than wiggly (hence
penalizing wigglyness). The Bayesian formulation allows us to be more
explicit about these beliefs \citep[section 5.8]{wood_generalized_2017-1}.
In general if we want to fit a model $y_{i}=s(x_{i})$ there is no
unique solution unless some restriction is put on the form of $s$
\citep{watson_smoothing_1984}. 

Looking at (\ref{eq:bayes-lik}), this says that $\bm{\beta}\sim N(\bm{0},\bm{S}_{\lambda}^{-})$,
where $\bm{S}_{\lambda}^{-}$ is the pseudoinverse of $\bm{S}_{\lambda}$.
Large penalty entries in $\mathbf{S}$ correspond to wiggly basis
functions (we want to penalize those more strongly) which, when inverted,
give small variances (our prior is that basis function's coefficient
is close to zero): this makes explicit our belief that smoother functions
are more likely than wiggly ones \citep{wood_confidence_2006}. 

Since often some of the elements of $\boldsymbol{\beta}$ are not
penalized (e.g., slope or intercept terms, which do not have derivatives),
this leads to improper priors as there are no constraints on the value
of the slope or intercept for those terms. In this case the pseudoinverse
of $\bm{S}_{\lambda}$ is required. Some basis-penalty smoothers do
lead to proper priors for all elements of $\boldsymbol{\beta}$ (e.g.,
the P-spline approach of \citealp{lang_bayesian_2004}) and generally
a proper prior can be found for any smooth by using the methods in
Section \ref{subsec:Term-selection,-proper}. Identifiability constraints
\citep[section 5.4.1]{wood_generalized_2017-1} that need to be imposed
on the model (e.g., that there is only one intercept in the model)
may also lead to proper priors \citep{marra_practical_2011}.

Various different basis function-penalty combinations available in
the literature express different priors on how we want our model terms
to behave. For example, cyclic smoothers give us terms which `match'
up to a set number of derivatives at the start/end of the data and
can be useful for temporal/seasonal effects. Many solutions have been
proposed to the issue of spatial smoothers in areas with complex coastlines
(e.g., \citealp{miller_finite_2014} and references therein). \citet{wood_soap_2008}
propose the soap film smoother, which simultaneously estimates a boundary
smooth while constraining values inside the boundary. These models
can be fitted using normal GAM machinery, since effects are generated
by transforms of the covariates (application of the basis functions)
and the prior (penalty) on the corresponding coefficients.

REML and ML are referred to as \textit{empirical Bayes} methods, as
when we take the random effects interpretation of $\boldsymbol{\beta}$,
we can think of this as a prior and our fit criterion assesses the
likelihood of the data given the implied prior on $\boldsymbol{\beta}$,
as in (\ref{eq:bayes-lik}). The `empirical' of the name indicates
that there is no prior for the smoothing parameters (see \citet{carlinEmpiricalBayesPresent2000}
for an overview of empirical Bayes methodology). Taking a fully Bayesian
approach, it is common to put a vague gamma prior on each element
of $\boldsymbol{\lambda}$ or uniform priors on their logarithm \citep{wood_just_2016}.
Specifying priors on smoothing/variance parameters can be tricky \citep{simpson_penalising_2017},
this is especially the case for smoothing parameters as the true values
of the smoothing parameter(s) could be infinite if the true smooth
is linear (right plot in Figure \ref{fig:EDF-lambda}). It can also
be hard to come-up with informed priors about smoothing parameters,
as we often do not have a direct interpretation of their values. 

When using splines we must also decide on knot placement/number and
basis complexity/dimension ($K$ in (\ref{eq:basis}); these are usually
linked). Since $\bm{S}_{\lambda}^{-}$ involves basis functions (or
at least their derivatives), the number of basis functions (and/or
number of knots) and knot placement will affect the posterior. Effects
of placement can be mitigated to some extent by over-specifying the
number of knots/basis functions and allow wigglyness to be dictated
by the smoothing parameter \citep[section 5.9]{pya_note_2016,wood_generalized_2017-1}.
Eigen-based approaches like thin-plate regression splines \citep{wood_thin_2003},
make placement data-based in cases where regular grids are computationally
taxing. Other related approaches include the use of triangulation-based
techniques to optimize placement based on data locations \citep{lindgren_explicit_2011}.

\subsection{Obtaining posteriors\protect\label{subsec:Posteriors}}

For a fully Bayesian (FB) approach we formulate a likelihood and attach
priors to the smoothing parameters $\boldsymbol{\lambda}$, as well
as the model coefficients $\bm{\beta}$. We could then use MCMC to
obtain a posterior. There are many software implementations which
can achieve this, so here I only list R packages specifically tailored
to GAMs: \texttt{mgcv::jagam}, which implements translation between
\texttt{mgcv} and \texttt{JAGS} \citep{wood_just_2016} or \texttt{brms}
\citep{b:urkner_brms_2017} which implements most \texttt{mgcv} models
in Stan \citep{carpenter_stan:_2017}. Dedicated software packages
such as \texttt{BayesX} \citep{brezger_bayesx_2005} can also be used.
If one wishes to avoid MCMC, integrated nested Laplace approximations
(INLA; often implemented via the \texttt{R-INLA} package) could be
used instead \citep{rue_approximate_2009,wood_simplified_2019}. Packages
that parameterize their multivariate normal distributions using precision
matrices rather than variances allow us to side-step the pseudoinversion
of the penalty discussed above. 

As discussed above, if we take an empirical Bayes (EB) view of the
world and do not put priors on $\boldsymbol{\lambda}$, we can still
obtain posteriors for $\bm{\beta}$, conditional on $\boldsymbol{\lambda}$.
For computational efficiency the Laplace approximation is often used
here. Both \texttt{R-INLA} (used in `empirical Bayes mode') and
\texttt{mgcv} use this approach.

Using either approach we can get to the posterior marginal for $\bm{\beta}$:
$\bm{\beta}\vert\bm{y},\bm{\lambda}\sim N(\hat{\bm{\beta}},\bm{V}_{\bm{\beta}})$
where for the Gaussian likelihood case $\bm{V}_{\bm{\beta}}=(\bm{X}^{\intercal}\bm{X}+\bm{S}_{\lambda})^{-1}\sigma^{2}$
and for the exponential family the expression is approximate and we
have $\bm{V}_{\bm{\beta}}=(\bm{X}^{\intercal}\bm{W}\bm{X}+\bm{S}_{\lambda})^{-1}\phi$,
where $\sigma^{2}$ is a variance parameter, $\phi$ is a scale parameter
and $\boldsymbol{W}$ is a weight matrix \citep[section 6.10]{wood_generalized_2017-1}.
For FB we can obtain a posterior for $\bm{\lambda}$ and an unconditional
posterior for $\bm{\beta}$. For EB we only have information conditional
on the value of the smoothing parameter(s). \citet{wood_smoothing_2016}
propose a correction to $\bm{V}_{\bm{\beta}}$ to account for the
uncertainty in the smoothing parameter(s) using a Taylor expansion
to approximate the extra variability in the smoothing parameter.

In practice we can fit our models using EB methods (such as using
REML/ML in \texttt{mgcv}) then sample from their posteriors. As noted,
this is as straightforward as plugging the mean coefficient estimates
and covariance matrix into a multivariate normal random number generator
when using a Gaussian likelihood, though in the exponential family
case one may have to use a Metropolis-Hastings sampler and proposing
from a \textit{t}-distribution to get reasonable results (such a sampler
can be accessed in \texttt{mgcv} using the \texttt{gam.mh} function).

\section{Some examples\protect\label{sec:Some-examples}}

The Bayesian results above lead to some useful applications. Here
I highlight a couple of the more commonly-used ones.  To illustrate
these techniques, data from the NOAA Alaska Fisheries Science Center's
groundfish assessment program (\url{https://www.afsc.noaa.gov/RACE/groundfish/survey_data/default.htm})
was used. The survey consists of summer bottom trawls at set of stations
from 1982 through to 2018 and are shown in Supplementary Figure 1.
Response was catch per unit effort (CPUE; measured as individuals
per hectare, effectively a density). Location (recorded as latitude/longitude
but projected for analysis), date, surface temperature, bottom temperature
(both recorded during the trawl, in degrees Celsius) and bathymetry
(recorded in metres) were available as covariates. See \citet{stevenson_bottom_2019}
(and references therein) for further details of the survey. The examples
below are not intended to be a serious analyses of the data. Data
were downloaded from the NOAA AFSC website and processed for this
analysis.

\subsection{Term selection\protect\label{subsec:Term-selection,-proper}}

We begin by fitting a model to the CPUE data for walleye pollock (\textit{Gadus
chalcogrammus}) in the Eastern Bering sea for 2010 only. The model
includes a bivariate smooth of location and then univariate smooths
of surface temperature, bottom temperature and bathymetry. Expected
CPUE was modelled as 
\begin{equation}
\mathbb{E}\left(\text{CPUE}_{i}\right)=\exp\left[\beta_{0}+s(x_{i},y_{i})+s(\text{Surface}_{i})+s(\text{Bottom}_{i})+s(\text{Depth}_{i})\right],\label{eq:cod-term-form}
\end{equation}
where $i$ indexes the station-years. CPUE was assumed to follow a
Tweedie distribution \citep[see e.g.,][for previous applications in fisheries]{shono_application_2008}
with a log link. To model CPUE, we may not need all of the covariates:
space ($x,y$), bottom depth ($\text{Depth}_{i}$), bottom temperature
($\text{Bottom}_{i}$) and surface temperature ($\text{Surface}_{i}$).
Rather than using hypothesis testing for term selection, here I apply
shrinkage/penalty-type methods to remove terms during model fitting,
effectively putting different priors on how to deal with the slope
and intercept in each smooth. Many approaches are possible \citep{marra_practical_2011}
but here I focus on two approaches implemented in \texttt{mgcv}.

As described in section \ref{subsec:Priors}, the prior placed on
$\bm{\beta}$ can be improper due to rank deficiency in $\mathbf{S}$.
This means that there are linear or intercept terms that are not penalized.
We refer to these terms as being \textit{in the nullspace} of the
penalty (the rest of the terms being referred to as the \textit{range
space}). Figure \ref{fig:EDF-lambda} illustrates this. We can make
our priors proper by simply adding an extra penalty term to the model
for the nullspace components of each term (the double penalty approach
of \citet{marra_practical_2011}). This is achieved by eigendecomposing
the penalty matrix, $\mathbf{S}=\mathbf{U}\bm{\Lambda}\mathbf{U}^{\intercal}$.
We can then form the additional penalty matrix $\mathbf{S}^{\ast}=\mathbf{U}^{\ast}\mathbf{U}^{\ast\intercal}$
where $\mathbf{U}^{*}$ is a matrix of eigenvectors corresponding
to the zero entries on the diagonal of $\bm{\Lambda}$. Our original
penalty $\mathbf{S}$ stays as-is, as the components in $\mathbf{S}^{*}$
do not have an effect (since their entries in $\boldsymbol{\Lambda}$
are (almost) zero). This approach is implemented as the \texttt{select=TRUE}
option in \texttt{mgcv::gam}, and includes one additional smoothing
parameter for each smooth term in the model, corresponding to each
term's nullspace. Alternatively one can form a basis where the terms
that lie in the nullspace have a shrinkage penalty applied to them
by simply adding a small value to their corresponding diagonal entries
of $\bm{\Lambda}$ so that the resulting penalty matrix is not rank-deficient
(the shrinkage approach of \citet{marra_practical_2011}; implemented
as the \texttt{cs} and \texttt{ts} bases in \texttt{mgcv}). One can
think of this as adding a ridge regression penalty to the nullspace
or, equivalently, as a regularization of the nullspace terms (see,
e.g., \citet{hooten_guide_2015} for further discussion of regularization
in ecology).

These two approaches lead to rather different interpretations of how
wigglyness should be penalized, or rather: the prior structure of
the smooths. The shrinkage approach assumes that the terms in the
nullspace should be penalized less than the other parts of the smooth
(since their contributions are small), so as the smoothing parameter
increases the model goes from very wiggly, to just the terms in the
nullspace (e.g., back to a linear model), to having no effect (estimated
as zero). This is appealing, as we can clearly see that increasing
the smoothing parameter (decreasing the variance scaling) results
in a less wiggly result, until the term is removed from the model.
The double penalty approach treats the null and range spaces separately
and makes no assumption about how much to smooth the nullspace components
relative to the other parts of the smooth. This means that the nullspace
components can be removed before the rest of the model, since there
is a smoothing parameter for each part.

\begin{figure}
\begin{knitrout}
\definecolor{shadecolor}{rgb}{0.969, 0.969, 0.969}\color{fgcolor}
\includegraphics[width=\maxwidth]{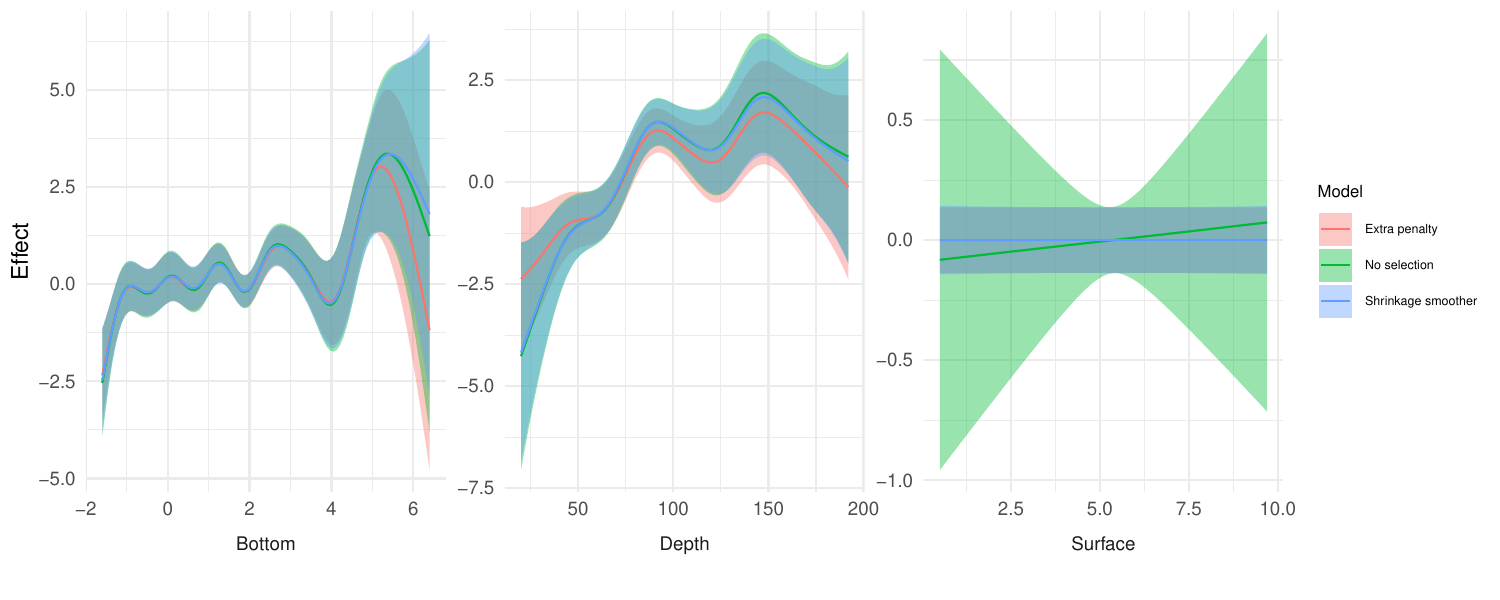} 
\end{knitrout}

\caption{Comparison using the shrinkage and double penalty approaches for term
selection, with no selection for reference. Models including bottom
temperature, depth and surface temperature (left to right) and spatial
terms (see Supplemental Figure 2) were fitted to
the walleye pollock CPUE data for 2010. Plots are on the linear predictor
(log) scale. Both the shrinkage and double penalty approaches remove
the surface temperature term (right), whereas the no selection method
(thin-plate regression splines) leave a linear term. Bottom temperature
uncertainty is estimated to be much smaller at the upper data range
for the extra penalty method. Other terms have minimal differences.
Note that confidence bands are generated including uncertainty in
the intercept for the top row.\protect\label{fig:shrinky}}
\end{figure}

We can fit (\ref{eq:cod-term-form}) in \texttt{mgcv} and see what
the differences are between the results using these different prior
specifications. Comparing the results from fitting all terms as thin-plate
regression splines (no selection), using the double penalty approach
and shrinkage revealed that the two term selection methods completely
removed the surface temperature term from the model. When no selection
method was used, the surface temperature term remained as a linear
term in the model (though it was not different from 0 according to
an F-test). Figure \ref{fig:shrinky} compares the resulting smooth
terms, though there are some minor differences the other smooths remain
the same between the three models (though this is not guaranteed in
general).

\subsection{Uncertainty around smooth terms\protect\label{subsec:Confidence-intervals}}

\begin{figure}
\begin{knitrout}
\definecolor{shadecolor}{rgb}{0.969, 0.969, 0.969}\color{fgcolor}
\includegraphics[width=\maxwidth]{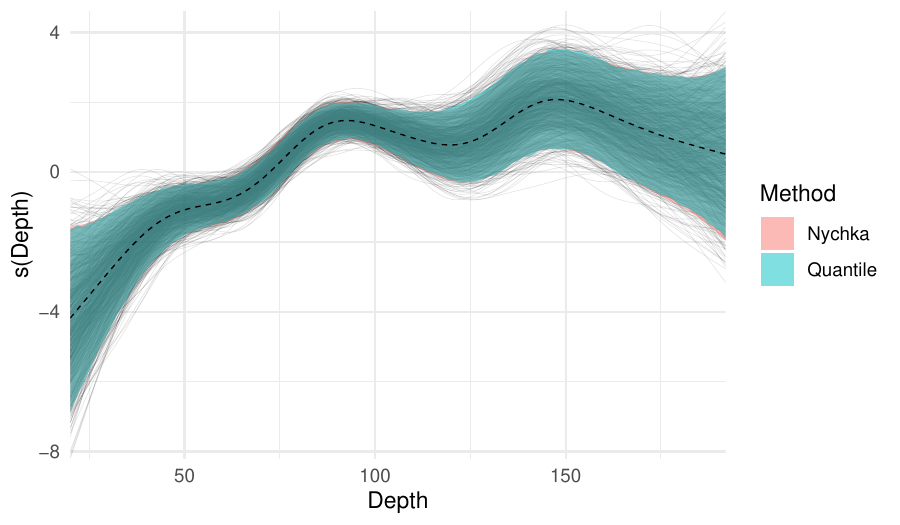} 
\end{knitrout}

\caption{Comparison of posterior samples and Nychka-type credible intervals
for the shrinkage model discussed in Section \ref{subsec:Term-selection,-proper}.
Dashed black line gives the mean smooth. 1000 posterior samples were
generated (black lines) using the algorithm given in Section \ref{subsec:Posterior-simulation},
95\% pointwise quantiles of the black lines are given by the green
ribbon. 95\% (Nychka-type) credible interval is also shown (red ribbon)
using the procedure in Section \ref{subsec:Confidence-intervals}.
\protect\label{fig:posterior-sampling}}
\end{figure}

From section \ref{subsec:Posteriors}, we could use the posterior
of $\bm{\beta}$ to generate possible parameters then use these to
generate possible smooths. From these simulated smooths, we could
then consider pointwise intervals over the range of the covariate
to build percentile confidence bands. Black lines in Figure \ref{fig:posterior-sampling}
shows 1000 posterior samples of the smooth of depth for the shrinkage
model in the previous section (black lines), and their 95\% quantiles
are the bounds of the blue band. We can take a shortcut and rather
than simulating, we know that each smooth can be written as a linear
combination ($s(x_{i})=\mathbf{X}_{i}\boldsymbol{\beta}$ for a model
with a single smooth in it). We can then use construct point-wise
credible intervals as $\hat{s}(x_{i})\pm z_{\alpha/2}\sqrt{v_{i}}$,
where $\hat{s}$ is our estimated smooth, $v_{i}$ is the variance
of the smooth at point $x_{i}$ and $z_{\alpha/2}$ is the usual appropriate
value from a normal CDF. Justification for these intervals was developed
in \citet{nychka_bayesian_1988} for normal responses and expanded
to the generalized case in \citet{marra_coverage_2012}. These intervals
have good frequentist across-the-function properties: that is, a 95\%
credible interval has close to 95\% coverage, when coverage is averaged
over the whole function. There may be over and under coverage at the
peaks and troughs of the function as we know less about the exact
turning points than we do about the function on the way to that turning
point (as by its nature we generally do not know if we have samples
at exactly the corresponding covariate value at the turning point).
The red band in Figure \ref{fig:posterior-sampling} shows these intervals.
Since these intervals have good coverage and tell us about the whole
function (by the across-the-function property), we can use them to
test the hypothesis $H_{0}:s(x)=0\;\forall x$ \textemdash whether
a term should be dropped from the model because it has no effect (the
$p$-values presented in output of \texttt{mgcv::summary}). See \citet{wood_p-values_2013}
for more detail on how $p$-values are calculated for this test. 

We use the posterior samples in Figure \ref{fig:posterior-sampling}
simply to calculate the blue band in the figure here but they can
be useful beyond this. Simulating from the posterior of smooth terms
(via simulation from the posterior of $\boldsymbol{\beta}$, conditional
$\boldsymbol{\lambda}$ or incorporating uncertainty via the approximation
described in Section \ref{subsec:Posteriors}) can potentially reveal
interesting properties of the fitted smooth which are not reflected
in the plotted bands.

\subsection{Posterior simulation/parametric bootstrap\protect\label{subsec:Posterior-simulation}}

Sometimes we want more than just uncertainty around individual terms
in the model, we want to know about uncertainty either in the model's
predictions or summary statistics generated from predictions. Since
we can simulate from the posterior of the model parameters, we can
use those parameters to calculate functions of the simulated parameters.
Calculating summary statistics on the results to obtain uncertainties
about those quantities. This is particularly powerful as it allows
us to calculate uncertainty about any function of the predictions
(including transformations which are non-linear, such as when applying
link functions, where this is necessary), avoiding potentially tricky
derivations needed to obtain analytical expressions for the variance
(see, for example, the derivations in the appendix of \citet{miller_estimating_2022}). 

\begin{figure}
\begin{knitrout}
\definecolor{shadecolor}{rgb}{0.969, 0.969, 0.969}\color{fgcolor}
\includegraphics[width=\maxwidth]{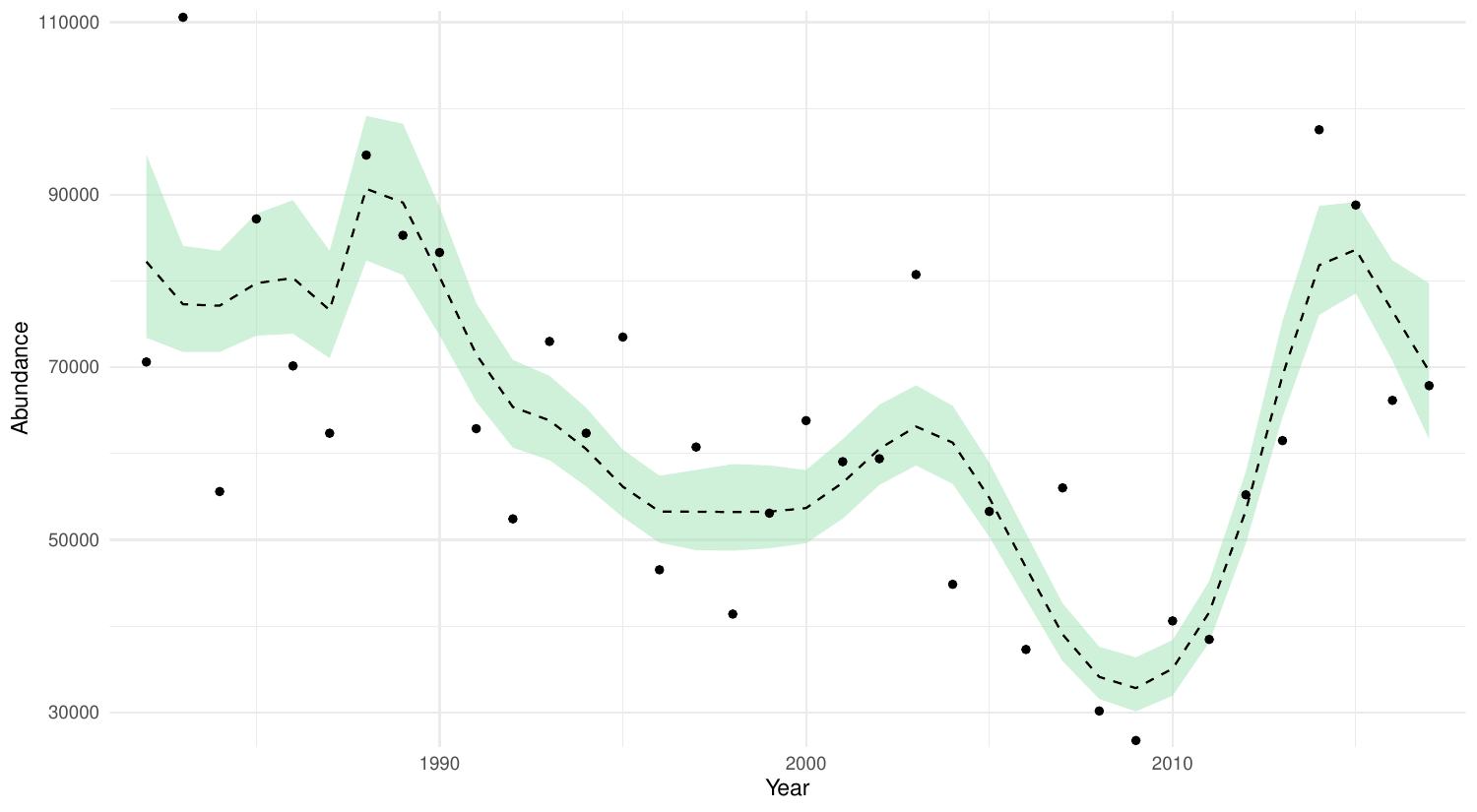} 
\end{knitrout}

\caption{Per-year estimates of total abundance at stations for the walleye
pollock data from a spatio-temporal model. Black dots indicate the
observed values (summed per year). Black dashed line shows the mean
of the samples from the posterior, summarised at the year level. The
green band shows a point-wise 95\% quantile interval. Note these are
not very smooth, as predictions are only made at the yearly level.
\protect\label{fig:walleye-temporal}}
\end{figure}

A general algorithm \citep[section 7.2.7]{wood_generalized_2017-1}
is as follows:
\begin{enumerate}
\item Let $B$ be the number of samples to generate.
\item Form $\mathbf{L}_{p}$, the matrix that maps the model covariates
to the linear predictor (the prediction equivalent of the design matrix).
\item For $b$ in $1,\ldots,B$:
\begin{enumerate}
\item Simulate $\bm{\beta}_{b}$ from the (approximate) posterior of $\bm{\beta}$.
\item Calculate the linear predictor $\bm{\eta}_{b}=\mathbf{L}_{p}\bm{\beta}_{b}$.
\item Apply the inverse link function, $g$, so $\bm{\mu}_{b}=g^{-1}(\bm{\eta}_{b})$.
\item Calculate and store the required summary of $\bm{\mu}_{b}$.
\end{enumerate}
\item Perform inference on the $B$ summaries (e.g., calculating empirical
variance, percentile intervals, etc).
\end{enumerate}
As an example of where we need to take summaries of non-linear functions
of the linear predictor, we can fit a spatio-temporal model to all
years (1982-2017) of walleye pollock CPUE data. Our model is then
$\mathbb{E}\left(\text{CPUE}_{i}\right)=\exp\left[s(x_{i},y_{i},t_{i})\right]$
(where $t_{i}$ indicates year). Now the smooth $s()$ is constructed
as a tensor product of a two dimensional thin-plate regression spline
smooth of $x$ and $y$, and a one dimensional cubic spline smooth
of $\text{Year}$. We want to obtain a time series of total predicted
abundance at the stations per year. So we need to predict at each
year and sum over the stations (i.e., space) at 3.(d) in the above
algorithm. Figure \ref{fig:walleye-temporal} shows the predictions
for the model, made by summing the predictions over space for each
year. For simplicity here abundance is calculated by summing over
the grid of all trawl station locations, it might be more appropriate
to sum over a finer spatial grid and since surveys are in the summer
only, abundance estimates were only made once per year (hence the
piecewise linear nature of Figure \ref{fig:walleye-temporal}). Increasing
the spatial or temporal resolution involves modifying $\mathbf{L}_{p}$
and recalculating steps 3.(c), 3.(d) and 4. (simulation from the posterior
does not have to be repeated).

\section{Discussion}

This article has highlighted the Bayesian interpretation of generalized
additive models (specifically as implemented in \texttt{mgcv}), which
are often thought of as a frequentist method. The article has emphasized
that ``GAM'' only describes a (very flexible) model structure and
that there are alternative ways to fit and interpret these models.
Taking a Bayesian interpretation gives us many ways in which these
links can exploited in practice for applied statistical work: they
are not only of mathematical interest.

Several topics have been excluded in this paper for reasons of brevity
and clarity, but curious readers may be interested in follow-on topics.
In (\ref{eq:gam}) we only consider the case where we are interested
in $\mathbb{E}(Y_{i})$ where $Y_{i}\sim\text{EF}(\mu_{i},\phi)$
but we need not restrict ourselves to these situations. There are
several additional distributions available within \texttt{mgcv} which
may be of use, including survival models (\texttt{cox.ph}), scaled
$t$-distributions (\texttt{scat}) and ordered categorical response
(\texttt{ocat}), as described in \citet{wood_smoothing_2016} (see
the \texttt{?family.mgcv} manual page for a full description of all
available distributions). We can also extend our models to GAMs for
location, shape and scale (GAMLSS; per \citet{rigby_generalized_2005}),
allowing for the specification of linear predictors for the shape
and scale parameters for many distributions including: normal (\texttt{gaulss}),
generalized extreme value (\texttt{gevlss}) and zero-inflated Poisson
(\texttt{ziplss}). Of some potential interest in ecology, are shape-constrained
splines which can be used to ensure that resulting smooths are, e.g.,
monotonically increasing/decreasing. These smoothers are implemented
in the \texttt{mgcv}-adjacent \texttt{scam} package \citep{pya_shape_2015}.

The Bayesian interpretations discussed here have been helpful to construct
more reasonable estimates of uncertainty (including smoothing parameter
uncertainty) and in order to understand how to construct confidence
intervals that have good coverage properties. In practice, \citet{miller_estimating_2022}
use the posterior simulation approach outlined here to obtain uncertainty
estimates for various aggregations of a complex spatio-temporal model
of fin whale abundance (including time series within and between years
and uncertainty maps). Since these uncertainty estimation schemes
are constructed in simulation-based approach, they can be significantly
easier to reason about and much easier to estimate uncertainty from
data subsets than traditional analytic estimates. Fitting the GAM
via REML/ML is fast (allowing for exploration), then uncertainty estimation
procedures are constructed by replacing appropriate steps in from
the simulation recipe given above.

Given the multivariate normal prior on the smoother parameters, $\hat{\boldsymbol{\beta}}$,
we can view a GAM as a Gaussian processes \citep[GP;][]{rasmussen_gaussian_2006}.
\citet{kimeldorf_george_s._correspondence_1970} give the general
theory for the theoretical link between stochastic processes (such
as GPs) and \citet{kent_link_1994} provide further details on links
between thin-plate regression splines and one specific type of GP:
kriging. Considering random effects as a specific type of basis function,
\citet{hefley_basis_2017} provide a more practical guide to this
equivalence, specifically with regard to highly structured spatiotemporal
data.

These links can surely be used further to develop other new methodology
and enhance our understanding of the models that we fit. This approach
has already been exploited to show that the stochastic partial differential
equation approach proposed by \citet{lindgren_explicit_2011} can
be viewed as a basis-penalty smoother and implemented in \texttt{mgcv}
\citep{miller_understanding_2019}. It is a shame that these conceptual
links have not been better recognized and exploited further; even
a very popular textbook \citep{ruppert_semiparametric_2003} describes
the mixed model representation of the GAM as a ``convenient fiction''.
Coming from the other direction, \citet{fahrmeir_bayesian_2010} expand
on the idea of Bayesian regularisation and its interpretation, deriving
corresponding priors for ridge regression, lasso, $L_{p}$ regularization,
elastic net, etc.

The \texttt{jagam} function (from \texttt{mgcv}) and the \texttt{brms} package
allow ecologists to quickly build models using familiar syntax very
similar to that for linear models, then transplant these into whatever
fully Bayesian computation system they prefer (see the recipe provided
by \citealp{miller_understanding_2019}). The models fitted in Section
\ref{sec:Some-examples} could be fitted in e.g., JAGS or Nimble,
using \texttt{jagam} to create necessary code. The main difference
between those models and the ones presented here would be the priors
on the smoothing parameters, which are not terribly interesting in
these cases. Where these ideas really shine are in allowing smooths
to be included as linear predictors for parameters in e.g., fully
Bayesian occupancy or mark-recapture models. A general strategy that
might be useful is using the GAM as a spatial distribution process
for the study species, but building more complex observation processes
(possibly from multiple data sources) in fully Bayesian framework
such as the one provided by Nimble. In this way, the complex spatial
structure is automatically generated and custom code is only required
to interface this part to the observation processes.

Moving beyond mere computational convenience and harnessing the broader
Bayesian framework implicit in this modelling strategy can help increase
understanding and synthesis, as well as providing further modelling
extensions within a familiar framework. 

\nocite{miller-data-1}
\nocite{miller-data-2}
\section*{Acknowledgements}

This work was partly funded by OPNAV N45 and the SURTASS LFA Settlement
Agreement, and being managed by the U.S. Navy's Living Marine Resources
program under Contract No. N39430-17-C-1982. Richard Glennie and Ian
Durbach provided extremely useful feedback on an early draft. John
Addy also provided extremely useful feedback later on. I also wish
to thank two anonymous reviewers who gave helpful comments on the
manuscript.

\section*{Conflict of Interest Statement}

No conflicts of interest.

\section*{Data Availability}

The Bering Sea data was downloaded from NOAA's Alaska Fisheries Science
Center at \url{https://www.afsc.noaa.gov/RACE/groundfish/survey_data/default.htm}
and was processed using scripts archived at \url{https://doi.org/10.5281/zenodo.14605076} \citep{miller-data-1}.
R code for the analyses presented here are available at \url{https://doi.org/10.5281/zenodo.14605014} \citep{miller-data-2}.

\printbibliography

\newpage

\section*{Supplementary Figures}

\setcounter{figure}{0} \renewcommand{\thefigure}{S\arabic{figure}} 

\begin{sidewaysfigure}
\begin{knitrout}
\definecolor{shadecolor}{rgb}{0.969, 0.969, 0.969}\color{fgcolor}
\includegraphics[width=\maxwidth]{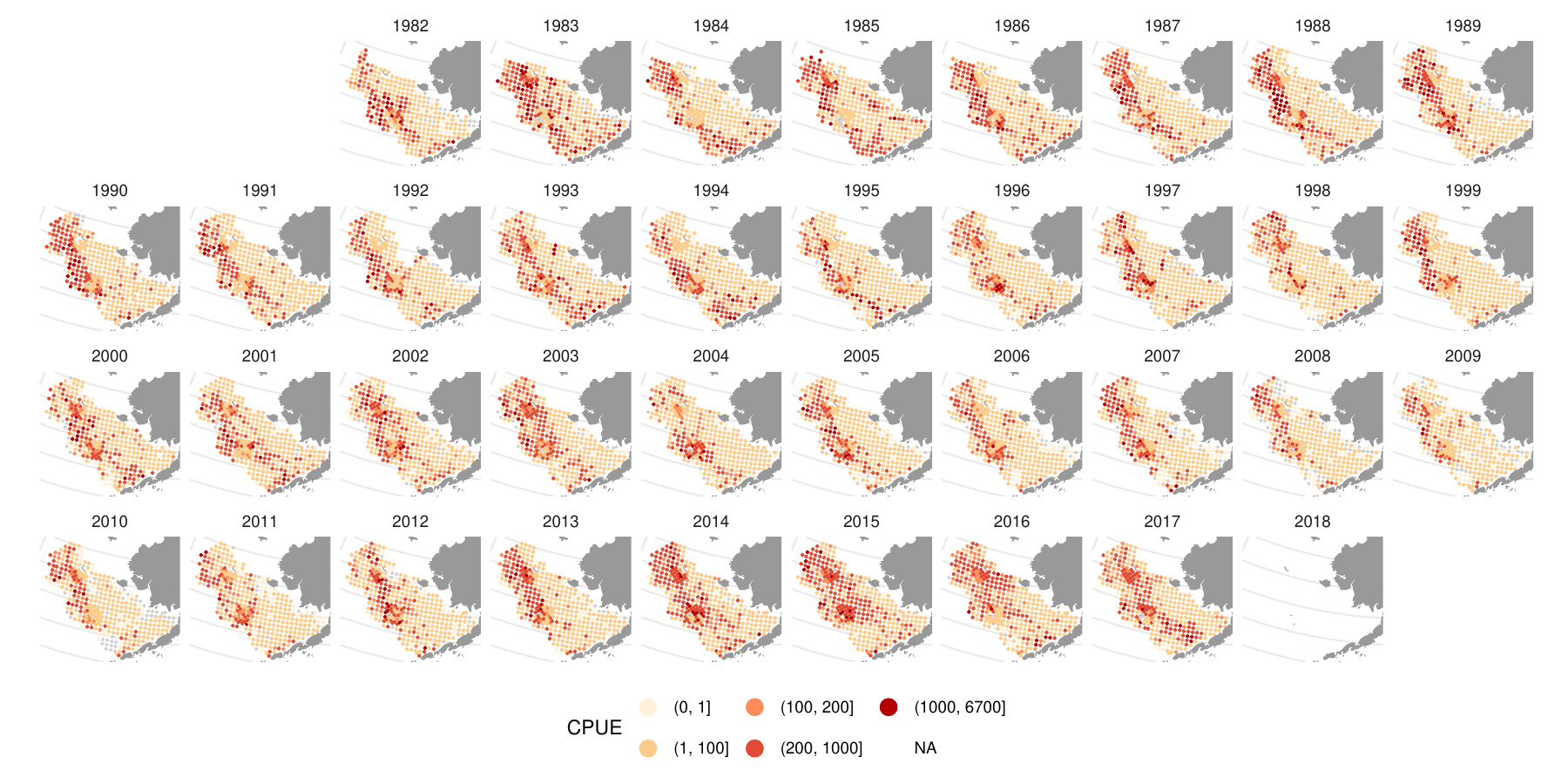} 
\end{knitrout}

\caption{Plot of the raw CPUE in space per year for walleye pollock. Empty
grey circles indicate sampling effort but no catch, colours indicate
catch per unit effort. Data from the NOAA Alaska Fisheries Science
Center's groundfish assessment program.\protect\label{fig:ak-fish-bois}}
\end{sidewaysfigure}

\begin{figure}
\begin{knitrout}
\definecolor{shadecolor}{rgb}{0.969, 0.969, 0.969}\color{fgcolor}
\includegraphics[width=\maxwidth]{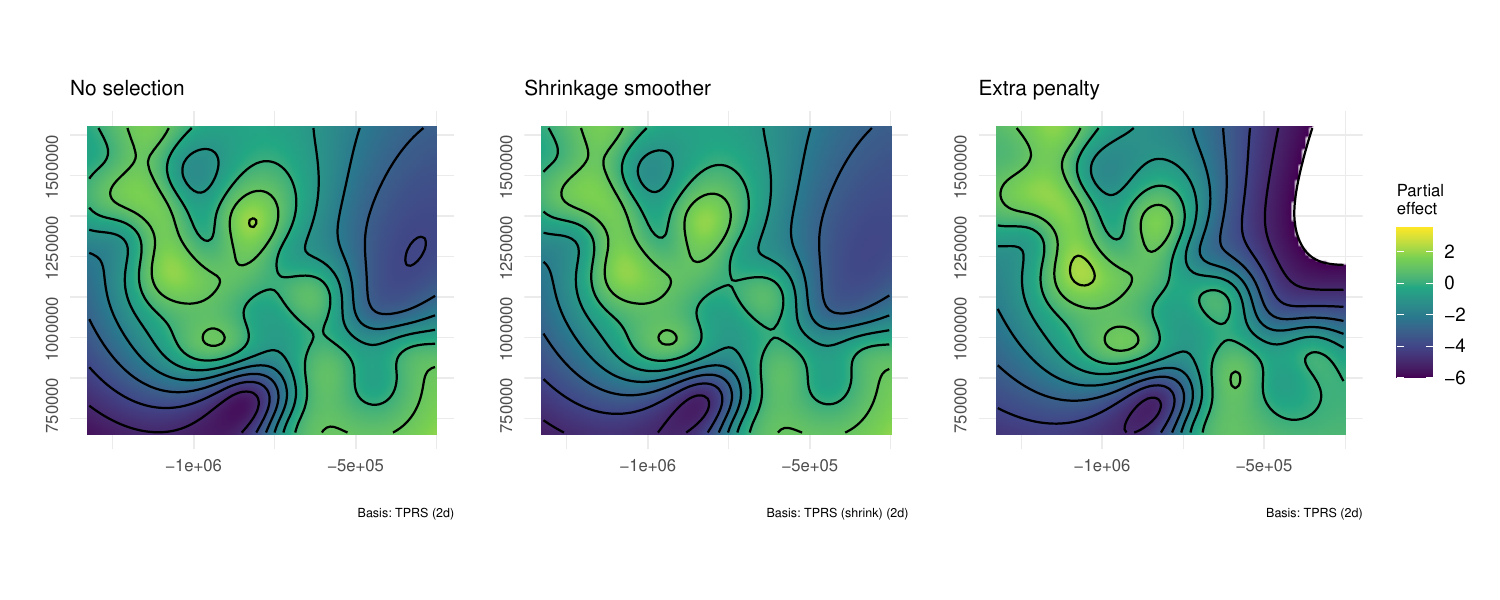} 
\end{knitrout}

\caption{Comparison using the shrinkage and double penalty approaches for term
selection, with no selection for reference. Models including smooths
of space (shown here), bottom temperature, depth and surface temperature
(see Figure 2) were fitted to the walleye pollock
CPUE data for 2010. Plots are on the linear predictor (log) scale.
\protect\label{fig:shrinky-1}}
\end{figure}

\end{document}